\documentclass[12pt]{article}

\usepackage{amsmath,amssymb,amsbsy,amsfonts,amsthm,latexsym,
                        amsopn,amstext,amsxtra,euscript,amscd,color}
\usepackage{color,xcolor}
\usepackage{latexsym}
\usepackage{cite}
\usepackage{amsfonts}
\usepackage{amsfonts,mathrsfs}
\usepackage{graphicx}
\usepackage{psfrag}
\usepackage{subfigure}
\usepackage{url}
\usepackage{stfloats}
\usepackage{amsmath}

\newtheorem{theorem}{Theorem}[section]
\newtheorem{lemma}[theorem]{Lemma}

\newtheorem{definition}[theorem]{Definition}

\newtheorem{example}{Example}

\newtheorem{conjecture}[theorem]{Conjecture}

\newcommand{\quash}[1]{}
\newcommand{\ceil}[1]{\left\lceil #1 \right\rceil}

\begin{document}

\title{On the $q$-Bentness of Boolean Functions}
\author{Zhixiong Chen$^{1}$, Ting Gu$^{2}$ and Andrew Klapper$^{3}$ \\
1. Provincial Key Laboratory of Applied Mathematics,\\
 Putian University, Putian, Fujian 351100, P.R. China\\
2. Elizabethtown College, PA 17022, USA\\
3. Dept.~of Computer Science,
University of Kentucky,\\ Lexington, KY
40506-0633, USA.\\
www.cs.uky.edu/$\sim$klapper.
}

\maketitle

\begin{abstract}
For each non-constant
$q$ in the set of $n$-variable Boolean functions, the {\em $q$-transform} of a Boolean function $f$
is related  to the Hamming distances from $f$ to the functions
obtainable from $q$ by nonsingular linear change of basis.
Klapper conjectured that no Boolean function exists with its $q$-transform coefficients equal to $\pm 2^{n/2}$ (such function is called $q$-bent).
In our early work, we only gave partial results to confirm this conjecture for small $n$. Here
we prove thoroughly that the conjecture is true by investigating the nonexistence of  the  partial difference sets in Abelian groups with special parameters.
We also introduce a new family of functions called almost $q$-bent functions, which are close to $q$-bentness.
\end{abstract}
\textbf{Keywords}.  Boolean function, Walsh-Hadamard transform, $q$-transform, $q$-bent function, partial difference set

\section{Introduction}
Since Rothaus invented bent Boolean functions in 1976 \cite{Rothaus} they have
been of significant interest to researchers.  Bent functions are those functions
that are as nonlinear as possible according to a certain theoretical bound.
More precisely, Parseval's identity says that the sum of the squares of the
Walsh-Hadamard coefficients of any Boolean function $f$ on $n$ bits is $2^{2n}$.
This implies that at least one affine function has distance no more than
$2^{n-1}-2^{n/2-1}$ from $f$.  Bent functions are those Boolean functions
for which the minimum distance to the affine functions is exactly
$2^{n-1}-2^{n/2-1}$.  They are important in cryptology because they are most
resistant to a linear approximation attack.

Recently Klapper generalized the Walsh-Hadamard transform to a class of
transforms, each transform associated with a non-constant Boolean function $q$
\cite{Kl16}.  This transform allows us to reason about the distance from
the set of functions obtained from $q$ by a linear change of coordinates.
The case when $q$ is linear is essentially the Walsh-Hadamard transform.
When $q$ is balanced, there is an analog of Parseval's identity and a bound
on the minimum distance from a Boolean function to the set of functions
obtained from $q$ by a linear change of coordinates.  This leads
naturally to a notion of {\em $q$-bent function}, those functions whose
distance to that set meets the theoretical bound.
More details on transforms and bent functions are discussed in Section \ref{sec-def}.

We conjectured that in fact, when $q$ is not affine, there are no $q$-bent
functions \cite{Kl16}.  In a later paper we proved that, when $q$ is not affine,
no function is both bent and $q$-bent \cite{KC}.
The main purpose of this paper is to settle (positively) the full conjecture.
This appears in Section \ref{sec-nonexist}.
The main technique used for this proof is the theory of {\em partial difference sets}.
In addition, in Section \ref{sec-almost} we introduce a notion of
{\em almost $q$-bentness} which measures how close a function is to
being $q$-bent.

\section{The Definition of $q$-bent Functions}\label{sec-def}

Let $n$ be a positive integer, let $V_n=\{0,1\}^n$, treated as row vectors, and let
${\cal B}_n=\{f:V_n\to \{0,1\}\}$, the set of Boolean functions of dimension $n$.
We refer the reader to Carlet's book chapter \cite{Ca10} and Cusick
and St\u{a}nic\u{a}'s monograph \cite{CS09} for background on
Boolean functions.

For $f,g \in {\cal B}_n$, let
\[
W(f,g) = \sum_{a\in V_n} (-1)^{f(a)+g(a)}\in \mathbb{Z}.
\]
Also let $d(f,g) = |\{a\in V_n: f(a)\ne g(a)\}|$, the Hamming distance
from $f$ to $g$.  Then
$$
W(f,g) = 2^n-2d(f,g).
$$
We denote by $wt(f)$ the \emph{weight} of $f\in {\cal B}_n$, which is
the cardinality of the \emph{support} of $f$, $\text{supp}(f)=\{a\in V_n : f(a)=1\}$.
Then we have $d(f,g)=wt(f+g)$, so
\[
W(f,g) = 2^n-2wt(f+g).
\]
For $a,v\in V_n$ let $v\cdot a$ denote the inner product of $a$ and $v$,
and let $g\in {\cal B}_n$ be the linear function $g(a)=v\cdot a$.  The
\emph{Walsh-Hadamard transform coefficient} of $f\in{\cal B}_n$ at $v$
is $W(f)(v)=W(f,g)$.  Parseval's identity says that
\[
\sum_{v\in V_n} (W(f)(v))^2 = 2^{2n}.
\]

A function $f \in {\cal B}_n$ with $|W(f)(v)|=2^{n/2}$ for all $v\in V_n$ is called a {\em bent function}.
For bent functions on $n$ bits to exist it is necessary that $n$ be even.
It is well known that bent functions exist for all even $n$.  For example,
a quadratic function is bent if and only if it has rank $n$.

Recently Klapper generalized the Walsh-Hadamard transform to a class of
transforms, each transform associated with a non-constant $q\in {\cal B}_n$
as follows \cite{Kl16}.
Let $GL_n=GL_n(\mathbb{F}_2)$ be the set of $n \times n$ invertible matrices
over $\mathbb{F}_2$, the finite field with two elements. Let $q_A$
denote the function $q_A(a) = q(aA)$ for $q\in {\cal B}_n$ and $A\in GL_n$.

\begin{definition}\label{q-trans}
Let $q\in {\cal B}_n$.  If $f\in {\cal B}_n$ and
$A\in GL_n$, then the {\em $q$-transform coefficient of $f$ at $A$}
is $W_q(f)(A) = W(f,q_A)$.  We also let
$W_q(f)(\mathbf{0}) = \sum_{a\in V_n} (-1)^{f(a)}\triangleq I_f$, where $\mathbf{0}$ denotes the zero matrix of $n\times n$.
\end{definition}
We note that $W_q(f)(A)$ is even for all $A\in GL_n\cup \{\mathbf{0}\}$.
Here $I_f$ denotes the imbalance of $f$. $I_f=0$ if and only if $f$ is balanced.

Klapper considered the statistical behavior
of the $q$-transform with respect to two probability distributions \cite{Kl16}.
For a random variable $X$ on $GL_n\cup \{\mathbf{0}\}$, let $E'[X]$
denote the expected value of $X$ with respect to the uniform
distribution on $GL_n$ and $E[X]$ denote the expected value of $X$ on $GL_n\cup \{\mathbf{0}\}$ with respect to
$\omega$, respectively, where $\omega$ is a probability
distribution on $GL_n\cup \{\mathbf{0}\}$ defined by
\[\omega(A) = \frac{1}{N+N/(2^n-1)}= \frac{2^n-1}{2^n N}\]
for $A\in GL_n$ and
\[\omega(\mathbf{0}) = \frac{N/(2^n-1)}{N+N/(2^n-1)}= \frac{1}{2^n},\]
where $N=|GL_n|$, which is the cardinality of the set $GL_n$.
Such choice of $\omega$ matches the standard distribution in the linear case.
Then, for balanced $q\in {\cal B}_n$, we have
$$
E'[W_q(f)(A)^2] = (2^{2n}-I_f^2)/(2^n-1),
$$
and so
$$
E[W_q(f)(A)^2] = 2^n.
$$
This is the generalization of Parseval's equation, which   leads to the notion of $q$-bent functions.

\begin{definition}(~\cite{Kl16}) \label{bent-def}
Let $q\in {\cal B}_n$ be balanced.  A function $f\in {\cal B}_n$ is $q$-bent if $|W_q(f)(A)|=2^{n/2}$ for all $A\in GL_n\cup \{\mathbf{0}\}$.
\end{definition}
Equivalently, the minimum distance from a $q$-bent function to
$\{q_A,q_A+1: A\in GL_n\cup \{\mathbf{0}\}\}$ is $2^{n-1}-2^{n/2-1}$.
As with bent functions, $q$-bent functions do not exist if $n$ is odd. If $q$ is linear, $f$ is  $q$-bent if and only if it is bent.
However, for non-affine balanced $q$, there are no known examples of $q$-bent functions and hence Klapper conjectured:

\begin{conjecture}(\cite{Kl16})\label{conj}
There are no $q$-bent functions for non-affine balanced functions  $q\in {\cal B}_n$.
 \end{conjecture}
It is true for $n=4$ by exhaustive search \cite{Kl16}. We also confirmed the conjecture if the rank of $q$ is less than
$3n/8$ or if $n$ is small  \cite{KC}, in which we have checked for $n\leq 22$ by searching the \emph{partial difference sets} in $V_n$, see the definition in Section \ref{sec-nonexist}.

In the following two sections, we show that Conjecture \ref{conj} is true for all
even $n$ by proving the nonexistence of partial difference sets with special
parameters in  Abelian groups.   Then we  introduce a new family of functions called
$\delta$-almost $q$-bent functions, where $\delta$ is a nonnegative integer.

\section{Nonexistence of $q$-bent functions}\label{sec-nonexist}
In this section we prove Conjecture \ref{conj}.

\begin{theorem}\label{main}
Let $n\geq 4$ be even. No $q$-bent functions exist for non-affine balanced
$q\in {\cal B}_n$.
\end{theorem}

To prove this theorem, we use the tool of partial difference sets
 due to the fact that any $q$-bent function has three-valued autocorrelation
for non-affine balanced $q\in {\cal B}_n$ if it exists \cite{KC}.

We now recall the notion of \emph{partial difference set} \cite{D-book,D,Ma84,Ma94,Ma97}. Let $(G,+)$ be an Abelian group\footnote{In
some papers $G$ is a group with addition \cite{D-book,D}, while in others
$G$ is a group with multiplication \cite{Ma84,Ma94,Ma97}.} of order $v$
with identity $0$ and let $D$ be a $k$-subset of $G$.
The \emph{different function}
of $D$ is defined by
\[
d_D(e)=|(D+e)\cap D|,~~~ e\in G,
\]
where $D+e=\{d+e: d\in D\}$ and $|Z|$ is the cardinality of a set $Z$.
 Then $D$ is called a $(v,k,\lambda,\mu)$ \emph{partial difference set} (briefly, PDS) in $G$ if
\[
d_D(e) =
\left\{
\begin{array}{ll}
\lambda,  &   \text{if }0\ne e\in D,\\
\mu, &   \text{if } 0\ne e\in G\setminus D.
\end{array}
\right.
\]
Furthermore, if $0\not\in D$ and $-D = D$, where $-D := \{-d : d \in D\}$, then
the partial difference set $D$ is called \emph{regular}.

A PDS $D$ is called \emph{trivial} if either $D\cup \{0\}$ or $G\setminus D\cup \{0\}$ is a subgroup of $G$. Otherwise $D$ is called \emph{nontrivial}.
For details on PDSs, the reader is referred to the work by Ma \cite{Ma84,Ma94,Ma97}.

By \cite[Proposition 3.1]{Ma84}, If $D$ is a partial difference set in an Abelian group $(G,+)$ with $-D=D$, so are
$D\cup \{0\}$, $D\setminus \{0\}$, $G\setminus D$, $(G\setminus D)\setminus \{0\}$, and
$(G\setminus D)\cup \{0\}$.

We have previously proved the following theorem.

\begin{theorem}\label{t1}(\cite{KC})
Let $n\geq 4$ be even.  Let $f\in {\cal B}_n$ be $q$-bent for non-affine balanced
$q\in {\cal B}_n$.
\begin{enumerate}
\item[(1)] If $0^n\not\in \mathrm{supp}(f)$, then $\mathrm{supp}(f)$ is a $(2^n,wt(f),\lambda,\mu)$ regular partial difference set in
$V_n$.
\item[(2)] If $0^n\in \mathrm{supp}(f)$, then $V_n\setminus \mathrm{supp}(f)$ is a  regular
$(2^n,2^n-wt(f),\lambda,\mu)$ partial difference set in $V_n$.
\end{enumerate}
In both cases above, $\lambda\neq \mu$.
\end{theorem}

For the remainder of this section we assumee that $D$ is a non-trivial
regular $(v,k,\lambda,\mu)$ PDS in an Abelian group $(G,+)$.  Let
\[\Delta=(\lambda-\mu)^2+4(k-\mu).
\]

The following lemma includes some known results of PDSs. For our
purpose, we will  put more restrictions on $|D|$.

\begin{lemma}\label{lemma2}
Let $G$ be an Abelian group of order $v$ and $D$ a $k$-subset of $G$ with $2\leq k<v-1$. If $D$ is a
non-trivial regular $(v,k,\lambda,\mu)$ PDS in $G$, then
\begin{enumerate}

\item[(1)]   $0\leq \lambda<k$, $1\leq \mu<k$;

\item[(2)]   $v,\Delta$, and $v^2/\Delta$ have the same prime divisors;

\item[(3)]   $(2k-(\lambda-\mu))^{2} \equiv 0 \pmod {\Delta}$; and

\item[(4)]   if $\Delta$ is a square, then $-\sqrt{\Delta}<\lambda-\mu<\sqrt{\Delta}-2$.
\end{enumerate}
\end{lemma}
Proof. By \cite[Lemma 1.3]{Ma97}. \qed

The notation $s^i\| t$ means that $s^i| t$ but $s^{i+1}\nmid t$ for integers $s,t$.

\begin{lemma}\label{prop1}
Let $G$ be an Abelian group of order $2^n$ and $D$ a $k$-subset of $G$.
If $D$ is a non-trivial regular $(2^n,k,\lambda,\mu)$ PDS in $G$ satisfying  $2^{\ell}\| k$ for some  $1\leq \ell<n$ and $\lambda \neq \mu$, we have
\begin{enumerate}
\item[(1)] $\Delta=2^{2m}$ for some positive integer $m$.
\item[(2)]  $2^{\ell +1}\|(\lambda-\mu)$.
\end{enumerate}
\end{lemma}
Proof.
(1): By Lemma \ref{lemma2}(2), we see that $\Delta$ is a power of two. By \cite[Theorem 5.2]{Ma84} $\Delta$ is a square, since otherwise
the cardinality of $G$ would be odd.

(2): By (1) in this lemma and Lemma \ref{lemma2}(3), we can write
\[
(2k-(\lambda-\mu))^2= \varepsilon^2 2^{2m}
\]
for some integer $\varepsilon$. Then we get that
\[
2k-(\lambda-\mu) = \pm \varepsilon 2^m.
\]
Under assumption, if we write $k=(2k_1+1)2^{\ell}$ for some integers
$k_1$ and $\ell$, we have
\[
\lambda-\mu =(2k_1+1)2^{\ell+1}\mp  \varepsilon 2^m.
\]
If $m\leq \ell+1$, we see that $2^m|(\lambda-\mu)$. But Lemma \ref{lemma2}(4) implies that $|\lambda-\mu|<\sqrt{\Delta}=2^m$
and so $\lambda-\mu=0$, a contradiction. Then we have $m>\ell +1$, which
implies that $2^{\ell +1}\|(\lambda-\mu)$.
 \qed

\begin{theorem}\label{thm3}
Let $G$ be an Abelian group of order $2^n$ with even $n$.
For $\lambda \neq \mu$, no non-trivial regular $(2^n,2^{n-1}\pm 2^{n/2-1},\lambda,\mu)$ PDS exists in $G$.
\end{theorem}
Proof.  Suppose that $D$ is a non-trivial regular $(2^n,k,\lambda,\mu)$ PDS in $G$ with  $k=2^{n-1}\pm 2^{n/2-1}$.

We consider the case where $k=2^{n-1}+ 2^{n/2-1}$ here.  The proof in the
case where $k=2^{n-1}- 2^{n/2-1}$ is similar\footnote{In this case, we
use $4\mu+ 2^{n/2+1}$ in place of $4\mu- 2^{n/2+1}$.}.
By Lemma \ref{prop1}(2), since $k=2^{n/2-1}(2^{n/2}+ 1)$, we suppose
 $\lambda-\mu=(2w+1)2^{n/2}$ for some integer $w$. Then we consider the equation
\begin{eqnarray*}
\Delta &=& (\lambda-\mu)^2+4(k-\mu)\\
&=& (2w+1)^2 2^{n}+ 4(2^{n-1}+ 2^{n/2-1}) - 4\mu\\
&=& (2w+1)^2 2^{n}+ 2^{n+1} - (4\mu- 2^{n/2+1}).
\end{eqnarray*}
 We note that, since $1\le \mu<k$ by Lemma \ref{lemma2}(1), we have
\[ - 2^{n/2+1} < 4\mu- 2^{n/2+1} < 2^{n+1}.
\]
 So only the conditions that
\[
\left\{
\begin{array}{l}
4\mu- 2^{n/2+1}=2^n\\
 (2w+1)^2=1+2+2^2+\cdots+2^{2m-n-1}
\end{array}
\right.
\]
can guarantee $\Delta=2^{2m}$ by  Lemma \ref{prop1}(1). However, $1+2+2^2+\cdots+2^{2m-n-1}=2^{2m-n}-1=(2^{(2m-n)/2}+1)(2^{(2m-n)/2}-1)$
 is a product of two consecutive odd numbers, which is not a square.
 This is a contradiction. \qed\\

\noindent
\textbf{Proof of Theorem \ref{main}.}
Now, if $f$ is $q$-bent for non-affine balanced $q\in {\cal B}_n$, we have $wt(f)=2^{n-1}\pm 2^{n/2-1}$. This is impossible by Theorems \ref{t1} and \ref{thm3}. \qed

\section{Almost $q$-bent functions}\label{sec-almost}

From above, $q$-bent functions don't exist for non-affine balanced $q\in {\cal B}_n$, we next ask how close we can get to $q$-bentness.
To that end we introduce a new class of Boolean function, which is slightly different from the one in \cite{KC}.

\begin{definition}\label{almostbent-def}
Let $q\in {\cal B}_n$ be balanced. A function $f\in {\cal B}_n$ is \emph{$\delta$-almost $q$-bent} if
\[
\left| |W_q(f)(A)|- 2^{n/2}\right|\leq \delta
\]
for all $A\in GL_n\cup\{\mathbf{0}\}$, where $\delta> 0$ is an integer.
The $q$-bentness of $f$ is the least integer $\delta$ such that
$f$ is $\delta$-almost $q$-bent.
The {\em bendability} of $q$ is the minimum $\delta$ such that there
exists a $\delta$-almost $q$-bent function.
\end{definition}
 Note that
Definition \ref{almostbent-def} is valid  for odd $n$ as well as even.
The bendability of a balanced $q$ is zero if and only if $q$ is a
non-zero affine function.
A $\delta$-almost $q$-bent function is $(\delta+1)$-almost $q$-bent,
but the converse is not in general true. Since $W_q(1+f)(A)=-W_q(f)(A)$
for all $A\in GL_n\cup\{\mathbf{0}\}$, we see that $f$ is $\delta$-almost
$q$-bent if and only if $1+f$ is $\delta$-almost $q$-bent.  If $f$ is balanced
then $W_q(f)(\mathbf{0}) = 0$.  Thus if $\delta < 2^{n/2}$, then a balanced $f$ is
not $\delta$-almost $q$-bent.

\begin{lemma}\label{wqf2}
Let $n>2$ be any integer (odd or even) and let $q\in {\cal B}_n$ be balanced.
For any $f\in {\cal B}_n$ and $A\in GL_n\cup \{\mathbf{0}\}$, we have
$W_q(f)(A)\equiv 0 \pmod 4$ if $wt(f)$ is even, and $W_q(f)(A)\equiv 2 \pmod 4$  if $wt(f)$ is odd.
\end{lemma}
Proof. The results follow from the facts that $W_q(f)(A)=2^n-2wt(f+q_A)$
for all $A\in GL_n$, that $W_q(f)(\mathbf{0})=2^n-2wt(f)$,
that $wt(q_A)=2^{n-1}$, and that $wt(f+q_A)$ equals the distance from
$f$ to $q_A$. \qed

\begin{theorem}\label{almostbent-evenwt}
Let $n\geq 4$ be an even integer  and let $q\in {\cal B}_n$ be a non-affine
balanced function.  Then
\begin{enumerate}
\item[(1).]  no $3$-almost $q$-bent function with even weight exists and
\item[(2).]  no $1$-almost $q$-bent function with odd weight exists.
\end{enumerate}
\end{theorem}
Proof. (1). Suppose that $f$ is $3$-almost $q$-bent with even weight. Since $3<2^{n/2}$,
$f$ is not balanced, hence is not affine.
We have for all $A\in GL_n\cup\{\mathbf{0}\}$
$$
2^{\frac{n}{2}} -3\leq  |W_q(f)(A)| \leq 2^{\frac{n}{2}} +3.
$$
By Lemma \ref{wqf2}, we see that $|W_q(f)(A)|= 2^{n/2}$ for all $A\in GL_n\cup\{\mathbf{0}\}$, this is impossible since no non-affine $q$-bent functions exist.

Statement (2) holds similarly. \qed\\

Suppose $n$ is even.
From Theorem \ref{almostbent-evenwt}, if a non-affine $f\in {\cal B}_n$ with even weight is $\delta$-almost $q$-bent for balanced non-affine $q\in {\cal B}_n$, then $\delta\geq 4$.
Similarly, if an $f\in {\cal B}_n$ with odd weight is $\delta$-almost $q$-bent for balanced non-affine $q\in {\cal B}_n$, then $\delta\geq 2$.


\begin{example} \label{ex-1}\em
Let $n=4$, $q(x)=x_1x_2+x_3$, and
$f(x)=x_1x_2+x_3x_4$ (so $wt(f)=6$).  By direct computation we see that
\[
|W_q(f)(A)|\in \{0, 4, 8\}, ~~~ A\in GL_n\cup\{\mathbf{0}\}.
\]
Thus $f$ is 4-almost $q$-bent (see a proof in the Appendix). We have also
checked that no 2-almost $q$-bent functions (with odd weight) exist
when $n=4$.
Thus the bendability of $q(x)=x_1x_2+x_3$ is 4 for $n=4$.

It is worth searching for 2 (or 4)-almost $q$-bent functions
for even $n>4$.\qed
\end{example}

\begin{theorem}\label{almostbent-oddwt}
Let $n\geq 3$ be an odd integer  and $q\in {\cal B}_n$ be a non-affine balanced function. Write $\omega=\lfloor 2^{n/2}\rfloor$, then
\begin{enumerate}
\item[(1).] no $3$-almost $q$-bent function with even weight exists, if $\omega\equiv 0$ or $3 \pmod 4$;
\item[(2).] no $2$-almost $q$-bent function with even weight exists, if $\omega\equiv 1$ or $2 \pmod 4$;
\item[(3).] no $3$-almost $q$-bent function with odd weight exists,  if $\omega\equiv 1$ or $2 \pmod 4$;
\item[(4).] no $2$-almost $q$-bent function with odd weight exists,  if $\omega\equiv 0$ or $3 \pmod 4$.
\end{enumerate}
\end{theorem}
Proof.
(1). Suppose that $f$ is $3$-almost $q$-bent with even weight. If $\omega\equiv 0 \pmod 4$  we find that
\[
\{a\in \mathbb{Z} : 2^{n/2} -3\leq  a \leq 2^{n/2} +3, 4|a\}=\{\omega\}.
\]
Lemma \ref{wqf2} tells us that $|W_q(f)(A)|=\omega$ for all $A\in GL_n\cup\{\mathbf{0}\}$.
However, $\omega<2^{n/2}$, which contradicts  the fact that
$E[W_q(f)(A)^2] = 2^n$.  So no $3$-almost $q$-bent function exists.
The case when $\omega  \equiv 3 \pmod 4$ can be proved similarly.  In this case $|W_q(f)(A)|=\omega+1$ for all $A\in GL_n\cup\{\mathbf{0}\}$.

(2). Now we suppose $\omega\equiv 2 \pmod 4$  and get
$$
\{a\in \mathbb{Z} : 2^{n/2} -2\leq  a \leq 2^{n/2} +2, 4|a\}=\{\omega+2\}.
$$
Since $\omega+2>2^{n/2}$, as above no $2$-almost $q$-bent functions exist. The case when $\omega \equiv 1 \pmod 4$ can be proved similarly.  In this case $|W_q(f)(A)|=\omega-1$ for all $A\in GL_n\cup\{\mathbf{0}\}$.

Statements (3) and (4) can be proved similarly.     \qed\\

From the proof of Theorem \ref{almostbent-oddwt}, if $f$ is $4$-almost $q$-bent,
we see that
$$
|W_q(f)(A)|\in\left\{
\begin{array}{ll}
\{\omega, \omega+4\}, & \mathrm{if}~~~ \omega\equiv 0 \pmod 4 \wedge~ wt(f) ~\mathrm{is ~ even},\\
\{\omega-3,\omega+1\}, &  \mathrm{if}~~~ \omega\equiv 3 \pmod 4 \wedge~ wt(f) ~\mathrm{is ~ even},\\
\{\omega-3, \omega+1\}, & \mathrm{if}~~~ \omega\equiv 1 \pmod 4 \wedge~ wt(f) ~\mathrm{is ~ odd},\\
\{\omega, \omega+4\}, &  \mathrm{if}~~~ \omega\equiv 2 \pmod 4 \wedge~ wt(f) ~\mathrm{is ~ odd}.
\end{array}
\right.
$$
If $f$ is $3$-almost $q$-bent,
we see that
$$
|W_q(f)(A)|\in\left\{
\begin{array}{ll}
\{\omega-2, \omega+2\}, & \mathrm{if}~~~ \omega\equiv 0 \pmod 4 \wedge~ wt(f) ~\mathrm{is ~ odd},\\
\{\omega-1,\omega+3\}, &  \mathrm{if}~~~ \omega\equiv 3 \pmod 4 \wedge~ wt(f) ~\mathrm{is ~ odd},\\
\{\omega-1, \omega+3\}, & \mathrm{if}~~~ \omega\equiv 1 \pmod 4 \wedge~ wt(f) ~\mathrm{is ~ even},\\
\{\omega-2, \omega+2\}, &  \mathrm{if}~~~ \omega\equiv 2 \pmod 4 \wedge~ wt(f) ~\mathrm{is ~ even}.
\end{array}
\right.
$$

\begin{example}\label{ex-2}\em
Let $n=3$ and let $q\in{\cal B}_3$ be a non-affine balanced function.
Then $q$ has degree 2, hence is linearly equivalent to $x_1x_2+x_3$ or to $x_1x_2+x_3+1$.  Thus we may assume $q(x)=x_1x_2+x_3$.
\begin{enumerate}
\item Let $f\in{\cal B}_3$ with $wt(f)=2$.
We have
$|W_q(f)(A)|\in \{0,4\}$ for  $ A\in GL_3\cup\{\mathbf{0}\}$.
So the $q$-bendability of $f$ is 3 (in this case $\omega=\lfloor 2^{n/2}\rfloor= 2 $).
The same holds if $wt(f)=6$
since $f$ is $\delta$-almost $q$-bent iff $f+1$ is.

\item
Let $f\in{\cal B}_3$ with $wt(f)\in \{1,3\}$.
We have
$|W_q(f)(A)|\in \{2,6\}$ for  $ A\in GL_3\cup\{\mathbf{0}\}$.
So the $q$-bendability of $f$ is 4. The same holds if $wt(f)\in \{5,7\}$
since $f$ is $\delta$-almost $q$-bent iff $f+1$ is.

\item
Let $f\in{\cal B}_3$ with $wt(f)=4$.  Then $f$ has degree 1 or 2,
hence is affine, is linearly equivalent to $x_1x_2+x_3$, or is linearly
equivalent to $x_1x_2+x_3+1$.  In the first case, we have
$\{|W_q(f)(A)|:A\in GL_3\cup\{\mathbf{0}\}\}= \{0,4\}$, so the
$q$-bendability of $f$ is 3.  In the second and third cases
$\{|W_q(f)(A)|:A\in GL_3\cup\{\mathbf{0}\}\}= \{0,4,8\}$, so the
$q$-bendability of $f$ is 6.
\end{enumerate}
It follows that the bendability of $q$ is 3.
\qed\\
\end{example}

\begin{example}\label{ex-3}\em
Let $n\ge 4$ and let $q(x)=x_1x_2+x_3\in {\cal B}_n$.  Suppose that $f\in {\cal B}_n$
is a quadratic form with rank $r$.   For any even $m>0$ let
$$B_m(x_1,\cdots,x_n) = x_1x_2 + x_3x_4 +\cdots + x_{m-1}x_m.$$
Recall that under linear equivalence either $f$
is equivalent to one the three types in Table \ref{table-classification}
(see \cite[Thm 6.30]{Lidl}).  Furthermore, the imbalance of $f$ is $2^{n-r/2}$
if it has Type I, 0 if it has Type II, and $-2^{n-r/2}$ if it has Type III.

\begin{table}[t]
\begin{center}
\begin{tabular}{ll|l|}
\cline{3-3}
\hline
\multicolumn{1}{|l|}{\bf Type I}& \multicolumn{1}{|l|}{($r$ even)} &
\multicolumn{1}{|l|}{$f(x)=B_r(x) $} \\
\hline \multicolumn{1}{|l|}{\bf Type II}&
\multicolumn{1}{|l|}{($r$ odd)}&
\multicolumn{1}{|l|}{$f(x)=B_{r-1}(x)+x_{r}$}  \\
\hline
\multicolumn{1}{|l|}{\bf Type III} &
\multicolumn{1}{|l|}{($r$ even)} &
\multicolumn{1}{|l|}{
$f(x)=B_{r}(x) +x_{r-1} + x_{r}$}  \\
\hline
\end{tabular}
\end{center}\caption{Classification of quadratic forms over $\mathbb F_2$}
\label{table-classification}
\end{table}

Thus $|W_q(f)(A)|$ depends only on the rank, $d_A$, of $f(x)+q_A(x)$.
We have $r-3\le d_A \le r+3$.  If $d_A$ is odd, then
$||W_q(f)(A)|-2^{n/2}| =2^{n/2}$, whereas if it is even, then
$||W_q(f)(A)|-2^{n/2}| =2^{n-d_A/2}-2^{n/2}$.  The latter number is greater
unless $d_A=n-1$.  Thus the bentness of $f$ is $\ceil{2^{n-d/2}-2^{n/2}}$,
where $d$ is the least even $d_A$.

Now suppose $f(x)=B_{r-1}(x)+x_r$, where $r$ is odd.  Choose $A$ so
$q_A(x) = x_{r-2}x_{r-1}+x_r$.  Then $f(x)+q_A(x)= B_{r-3}(x)$, so $f$ is
$\ceil{2^{n-(r-3)/2}-2^{n/2}}$-almost $q$-bent.

On the other hand, suppose $r\ge 4$ is even and suppose $f(x)=B_{r}(x)$.
Choose $A$ so $q_A(x) = x_{r-1}x_{r}+x_{r-2}$.  Then
$f(x)+q_A(x)= B_{r-4}(x)+x_{r-3}(x_{r-2}+x_{r-3})$.  This has rank
$r-2$, so $f$ is $\ceil{2^{n-(r-2)/2}-2^{n/2}}$-almost $q$-bent.
If $r=2$ and $f(x)=B_{r}(x)$, then $f$ is $\ceil{2^{n-1}-2^{n/2}}$-almost
$q$-bent.  The same results hold if $f$ has Type III.

It follows that the minimal $q$-bentness among quadratic Boolean functions $f$
is $\ceil{2^{n-(n-2)/2}-2^{n/2}}=\ceil{2^{n/2}}$ when $n$ is even
(with $r=n$).  It is $\ceil{2^{(n+3)/2}-2^{n/2}}=\ceil{2^{n/2}(2^{3/2}-1)}$
when $n$ is odd (with $r=n-1$ or $r=n$).

Similar analysis can be applied when $q$ has odd rank $k>3$.  In this case
$r-k\le d_A \le r+k$.  Again, the $q$-bentness of $f$ is determined by the
least possible even $d_A$.  We omit the details.
\qed\\
\end{example}

\section{Conclusions}

We have shown that there are no non-affine Boolean functions, the minimum distance from which to
the functions obtainable
from $q\in {\cal B}_n$ ($n$ even, $q$ balanced and not affine) by a linear change of coordinates can achieve $2^{n-1}-2^{n/2-1}$.
This result answers a
conjecture proposed by Klapper \cite{Kl16}. That is to say,  $q$-bent functions are nothing but the bent functions when $q$ is linear.

We also have given a new family of functions named $\delta$-almost $q$-bent functions.
We have illustrated $3$-almost and $4$-almost $q$-bent functions for $n=3$ and $4$-almost $q$-bent functions for $n=4$ with some examples.
Such functions are the almost $q$-bent functions of least $\delta$.
however, we cannot give an example of $2$-almost $q$-bent functions for even $n$. We leave it open. We note that if $f$ is $2$-almost $q$-bent for even $n$,  then $|W_q(f)(A)|\in \{2^{n/2}-2, 2^{n/2}+2\}$.
It is also interesting to search
for  $\delta$-almost $q$-bent functions of least $\delta$ for $n>4$.

As mentioned in our earlier paper \cite{KC}, there is a coding theoretic
interpretation of our main result.  We can interpret a Boolean function
$f \in {\mathcal B}_n$ as a vector of dimension $2^n$, indexed by $V_n$.
Then distance between functions corresponds to Hamming distance.
For fixed a non-affine $q\in {\mathcal B}_n$, we can consider the nonlinear code
${\mathcal C}_q=\{q_A: A\in GL_n\cup \{\mathbf{0}\}\}
\cup \{1+q_A: A\in GL_n\cup \{\mathbf{0}\}\}$.  Then the fact
that no $f \in {\mathcal B}_n$ is $q$-bent implies that for some
$A\in GL_n\cup \{\mathbf{0}\}$ we have $dist(f,q_A)<2^{n-1}-2^{n/2-1}$.
Equivalently, the covering radius of ${\mathcal C}_q$ is less than
$2^{n-1}-2^{n/2-1}$.

\section*{Acknowledgments}
\indent
The authors wish to thank prof. Cunsheng Ding for some suggestions on the theory of difference sets.

The work was partially supported by the National Natural Science
Foundation of China under grant No.~61772292.

A. Klapper was partially supported by the National Science Foundation
under Grant No.~CNS-1420227. Any opinions, findings, and conclusions or recommendations expressed in this material are those of the authors and do not necessarily reflect the views of the National Science Foundation.

\appendix

\section{Example \ref{ex-1} Revisited}
In this appendix we give a proof for Example \ref{ex-1}.

We write $\mathrm{supp}(f)=\{\alpha_i : 1\leq i\leq 6\}$ with
$$
\alpha_1=1100, ~ \alpha_2=1011, ~ \alpha_3=0111, ~ \alpha_4=0011,~ \alpha_5=1101, ~ \alpha_6=1110.
$$
We check that $\alpha_1+\alpha_2+\alpha_3=0000$ and  $\alpha_4+\alpha_5+\alpha_6=0000.$ However, in the set $\mathrm{supp}(q)$,
we only have $1100+1010+0110=0000$ and among vectors left there are no three vectors such that they are linearly dependent.
Similar result holds in the set $V_4\setminus (\mathrm{supp}(q)\cup\{0000\})$. This means that there are no $A\in GL_4$ such that
$\mathrm{supp}(f_A)\subseteq \mathrm{supp}(q)$ or $\mathrm{supp}(f_A)\subseteq V_4\setminus (\mathrm{supp}(q)\cup\{0000\})$. In other words,
we always have
$$
\mathrm{supp}(f_A)\cap \mathrm{supp}(q)\neq \emptyset$$
and
$$
\mathrm{supp}(f_A)\cap V_4\setminus (\mathrm{supp}(q)\cup\{0000\}))\neq \emptyset
$$
for all $A\in GL_4$. Hence we get
$$
wt(f_A+q)\in \{4,6,8,10,12\},
$$
from which we derive $W_q(f)(A)\in \{0,\pm 4, \pm8\}$ for all $A\in GL_4$. \qed

\end{document}